\documentclass{llncs}
\usepackage[title]{appendix}
\makeatletter
\newcommand{\@chapapp}{\relax}%
\makeatother
\usepackage{amsfonts}
\usepackage{amsmath}
\usepackage{tikz}
\usepackage{tkz-euclide}
\usepackage{ifthen}
\usepackage{myCustomMathsSymbols}
\usepackage{xcolor}
\usepackage{hyperref}
\usepackage[normalem]{ulem}
\usepackage{caption}
\usepackage{lscape}
\usepackage{soul}
\usepackage{pgfplots}
\usepackage{subcaption}
\usepackage{stackengine}
\pgfplotsset{compat = newest}

\useunder{\uline}{\ul}{}
\def\id{{\mathbf{1}}}
\def\white#1{\textcolor{white}{#1}}
\newcommand\scalemath[2]{\scalebox{#1}{\mbox{\ensuremath{\displaystyle #2}}}}
\input{Qcircuit}	

\begin{document}

\title{Grover's oracle for the Shortest Vector Problem and its application in hybrid classical-quantum solvers}

\author{Milo\v{s} Prokop\inst{1,2}, Petros Wallden\inst{1}, David Joseph\inst{2}}
\institute{School of Informatics, University of Edinburgh \\ \email{m.prokop@sms.ed.ac.uk}
\\ 
\email{petros.wallden@ed.ac.uk}
\and
SandboxAQ \\
\email{david.joseph@sandboxaq.com}}

\markboth{Journal of \LaTeX\ Class Files,~Vol.~14, No.~8, August~2015}%
{Shell \MakeLowercase{\textit{et al.}}: Bare Demo of IEEEtran.cls for IEEE Journals}
\maketitle

\begin{abstract}
Finding the shortest vector in a lattice is a problem that is believed to be hard both for classical and quantum computers. Many major post-quantum secure cryptosystems base their security on the hardness of the Shortest Vector Problem (SVP) \cite{NIST_talk}. Finding the best classical, quantum or hybrid classical-quantum algorithms for SVP is necessary to select cryptosystem parameters that offer sufficient level of security. 
Grover's search quantum algorithm provides a generic quadratic speed-up, given access to an oracle implementing some function which describes when a solution is found. 
In this paper we provide concrete implementation of such an oracle for the SVP. We define the circuit, and evaluate costs in terms of number of qubits, number of gates, depth and T-quantum cost. We then analyze how to combine Grover's quantum search for small SVP instances with state-of-the-art classical solvers that use well known algorithms, such as the BKZ \cite{schnorr1994lattice}, where the former is used as a subroutine. This could enable solving larger instances of SVP with higher probability than classical state-of-the-art records, but still very far from posing any threat to cryptosystems being considered for standardization. Depending on the technology available, there is a spectrum of trade-offs in creating this combination. 

\end{abstract}

\section{Introduction}

The most powerful known quantum algorithm, of Shor \cite{shorsAlgo}, is renowned for its implications for public key cryptography. The ability to factor integers efficiently fatally undermines the security of cryptosystems such as RSA \cite{rivest1978method}, Elliptic Curve Cryptography \cite{blake1999elliptic}, and Diffie-Hellman key exchange \cite{diffie2022new} which together form a basis for authentication and secure key exchange over the internet.

While Shor's algorithm has been known for around 30 years, the recent surge of progress in quantum computing hardware and engineering has motivated a number of works analyzing the resource requirements of implementing such algorithms. 
There are often different circuits and subroutines that perform the same operation in different ways, variously requiring more auxiliary qubits or higher depth, and the impact of using different subroutines must be analyzed.
The effect of error correction and noisy qubits is also considered. The quantum computing hardware race is still wide open, with seven distinct technological approaches, so it is impossible to even know which platform (and hence with which optimizations) the first crypto-cracking algorithms will be run on.
	
In response to the quantum threat, the cryptography community has devised a new set of cryptographic primitives known as post-quantum cryptography (PQC) which we expect to remain resilient against quantum attacks. However it is generally acknowledged that going forward, parameter sets for all cryptographic algorithms will have to take into account quantum attacks (usually variations of Grover search) and thus designs aiming for a set security level must work backwards to understand the quantum circuit complexity that an attacker would require to break the cryptosystem.
	
One of the key families of PQC is based on lattice problems (versus integer factorization which is used presently). Three of four algorithms recently announced to be standardized by the National Institute for Standards and Technology (NIST) are based on lattice constructions, and one of the hard problems that must remain hard in order for lattice cryptography to remain secure is called the Shortest Vector Problem (SVP). The security of SVP has been widely investigated from a classical perspective, but quantum investigations are less advanced. On the more theoretical end, quantum tree search algorithms have been applied to enumeration techniques, while less success has been achieved applying quantum methods to sieving. On the other hand, `full-fat' quantum SVP algorithms have been proposed based on encoding SVP to the ground state of a  Hamiltonian, and this has been investigated in adiabatic settings \cite{joseph2020not}, on quantum annealers \cite{vqaSVPDavid}, and in the gate model \cite{vqaSvpMilos,zhu2022iterative}. The drawback of some of these quantum-native approaches is that time-complexity results are far scarcer for this creed of quantum algorithms and is difficult to quantify the concrete cost for instances far from those that can realistically implemented or simulated currently. 

As stressed above, many quantum attacks use different types of search subroutines that are all based on Grover's search algorithm.  Since Grover's algorithm, fully or as part of larger hybrid algorithms, appears in all these cases, it is important to analyse the exact cost of implementing Grover's algorithm for SVP, to quantify the concrete cost and enable calculating the security level that post-quantum cryptosystems offer. Implementing Grover's search for SVP efficiently, can form a module that could potentially be used in a variety of existing and new attacks. Specifically, building on the Hamiltonian approaches to SVP, 
one can reduce the concrete resource cost. 

\subsection{Our contributions}

In this contribution we aim to implement efficiently 
and modularly the Grover's oracle for SVP and then estimate its costs and some direct implications it has within larger algorithms.

\begin{itemize}
    \item We implement Grover's oracle for the Shortest Vector Problem, giving a detailed analysis of the circuit design.
    \item We estimate the resources required for the oracle, including the space and time complexities, the number of gates as well as the number of $T$-gates.    
    \item Given our oracle implementation, we estimate the resources required for running the full Grover's algorithm in order to solve SVP for different lattice dimensions. 
    \item We discuss the implications of our analysis for improving the performance of the classical, state-of-the-art, BKZ algorithm for solving SVP. Specifically we use Grover's search as subroutine within BKZ and discuss the improvements one could get in terms of accuracy or running time for solving as large dimension as possible of the SVP and comment on the relevance for existing postquantum secure cryptosystems. 
\end{itemize}
Here it is important to note that in our calculations we assume perfect (noise-less) qubits. In implementing those algorithm in realistic noisy setting, overheads due to quantum error correction would be needed. These overheads depend on many things, including on the quantum error correcting codes used, noise level and other characteristics of the hardware, etc. Our results can easily be re-interpreted for all these settings by considering the gates as ``fault-tolerant'' gates (and their corresponding times), and similarly the number of qubits as the number of ``logical'' qubits. To simplify such uses of our results, we also provide the $T$-gate count, since those gates would have different cost when implemented fault-tolerantly.

\subsection{Related Work}
Many quantum algorithms that include an oracle have been proposed, e.g. Grover's algorithm \cite{groversAlgo}, Deutsch-Jozsa algorithm \cite{deutschJozsa}. Shor's algorithm \cite{shorsAlgo}, or for the direct application to solving the SVP problem, the Montanaro's backtracking algorithm \cite{montanaroBacktracking}. This oracle is a black-box function that can be easily described by an analytical formula while a practical circuit implementation is not being considered. This facilitates high-level analysis of the algorithms omitting the essential practical implementation details. Several works have discussed approaches to implement the quantum oracles. \cite{Zhao_2022} discusses small-scale implementation of the quantum oracles for the first three aforementioned algorithms, \cite{Henderson2023AutomatedQO} proposes an approach for automatic synthesis of the quantum oracles for Grover's and Shor's algorithms, \cite{oracleAmplitudeAmplification} proposes a circuit for Grover's algorithm to find elements less than a certain value in an unstructured database. \cite{aesResourceEstimation} implements Grover's oracles different key sizes of Advanced Encryption Standard (AES). They decompose the circuits into Clifford+T set and argue about qubit and depth requirements.

In the security arena, researchers have begun to investigate constructions for symmetric cryptography. To evaluate the complexity of breaking the Advances Encryption Standard (AES), 
resource estimates were calculated for building an oracle to perform exhaustive key search \cite{aesResourceEstimation}. For hash functions, oracle constructions revealed that due to the varying dependencies on addition or multiplication, SHA2 may remain harder for quantum computers to break than its successor, SHA3 \cite{preston2022applying}. For SVP specifically, constructions have shown how to encode the problem as a Hamiltonian, and analyzed its performance in algorithms ranging from adiabatic quantum computation \cite{joseph2020not} to variational quantum eigensolvers \cite{joseph2022quantum,zhu2022iterative} and annealers \cite{vqaSVPDavid}.
More recently, one independent work has found lower bounds on circuit depth while searching for short vectors via enumeration with extreme pruning \cite{bindel2023quantum}. Another gives a concrete implementation of a circuit for enumarion using Montanaro's quantum tree backtracking algorithm \cite{bai2023concrete}.

\subsection{Structure of the paper}

In Section \ref{sec:preliminaries} we give the necessary background, namely we introduce the Shortest Vector Problem,  Grover's search algorithm and the BKZ algorithm for solving SVP. In Section \ref{sec:figsOfMerit} we define the figures of merit that we will use when evaluating the resources required for (sub)algorithms and in Section \ref{sec:circuitDesign} we give an implementation of Grover's oracle suited for the SVP. Section \ref{sec:resourceEstim} we analyzes resource requirements of the single oracle circuit and Section \ref{sec:solvingSVPusingGrover} further discusses the overall resource requirements for the full Grover's search run for different lattice dimensions up to one hundred. In Section \ref{sec:groversImprovement} we analyze how using Grover to solve SVP for relatively smaller dimensions, can also be important, by using it as a subroutine to improve the classical BKZ algorithm \cite{bkz20}, and give the potential improvements. We summarize and conclude in Section \ref{sec:conclusions}.

\section{Preliminaries}\label{sec:preliminaries}

\subsection{Lattices}
A mathematical lattice is a repeating pattern of points in $\mathbb{R}^n$. It can be described by a single basis vector $\mathbf{b}_i$ in each dimension, so $n$ linearly independent vectors. The basis vectors taken together are known as a basis $\mathbf{B} = (\mathbf{b}_1, \hdots, \mathbf{b}_n)$. There are infinitely many bases that can be used to describe a single lattice, and each is related to every other by a unimodular transformation $\mathbf{B}' = \mathbf{U}\mathbf{B}$. Some bases contain short vectors which are are close to orthogonal, said to be good, and others contain only very long vectors. Finding short bases can be used as an intermediate step to find short vectors, and vice versa. Of a particular importance will be for us \textbf{Gaussian Heuristics} which is an estimate on the length of the shortest non-zero vector of the full-rank lattice $\mathcal{L}$ $$\text{gh}(\mathcal{L})=\sqrt{n/2\pi e}\text{Vol}(\mathcal{L})^{1/n}$$

and \textbf{Gaussian Orhogonalization} of lattices. Denote the orthogonal projection $\pi_i \colon \mathbb{R}^n \rightarrow (\mathbf{b}_1, \hdots, \mathbf{b}_{i-1})^\perp$. Write $\pi_i(\mathbf{B}_{[i : j]}) = (\pi_i(\mathbf{b}_1), \hdots, \pi_i(\mathbf{b}_{j-1}))$ and let the corresponding lattice be known by $\pi_i(\mathcal{L}_{[i : j)})$.
Let $\mathbf{B}^*=(\mathbf{b}_1^*, \hdots, \mathbf{b}_n^*)$ denote the Gram-Schmidt orthogonalization of the basis. Then $\pi_i(\mathbf{b}_i) = \mathbf{b}_i^*$.

\subsection{The Shortest Vector Problem (SVP)}
One of the fundamental hard problems in the study of mathematical lattices is how to find the shortest vector, known as the shortest vector problem (SVP). 
\begin{definition}
Given a basis $\mathbf{B}$ that describes a lattice $\mathcal{L}(\mathbf{B})$, find the shortest vector $\mathbf{v} \in \mathcal{L}$, which has length $\| \mathbf{v} \| = \lambda_1(\mathcal{L})$. 
\end{definition}
A more relaxed variant of SVP is called $\gamma$-approximate SVP, or SVP$_\gamma$. This requires the solver to find a vector $\mathbf{v} \in \mathcal{L}$ such that $\| \mathbf{v} \| \leq \gamma \cdot \lambda_1$ where $\gamma=\text{poly}(n)$. It is this more relaxed version that seek solutions for in this work.
We note that technically, the zero vector $\mathbf{0}$ is a vector in all lattices by definition. However this trivial solution is not a permissible SVP solution. This is an important when constructing quantum Hamiltonians describing SVP solutions in the following.

\subsection{SVP algorithms}
There are two approaches to solving SVP. The first is sieving, where one samples many vectors from a somewhat short distribution and iteratively combines them until probabilistically reaching a short vector. The other is enumeration where one effectively counts all the vectors inside a ball of a determined size deterministically.

Classically, the cost of the enumeration subroutine is $2^{O(n \log n)}$ \footnote{Unless a base of a logarithm is specifically stated, in this paper we consider $log(\cdot)$ to be a logarithm with base 2.} on an $n$-dimensional lattice, which for our BKZ purposes becomes $\beta$. Lattice sieving runs faster, taking $2^{0.292\beta + o(\beta)}$ time, though it requires exponential space. Typically enumeration outperforms sieving for small problem sizes until a crossover point around at around dimension $70$ \cite{albrecht2019general}. That is to say, despite better asymptotic performance, enumeration is still a better choice for some practical choices of $n$.

\subsection{Grover's algorithm}\label{sec:prelimGroversAlgo}
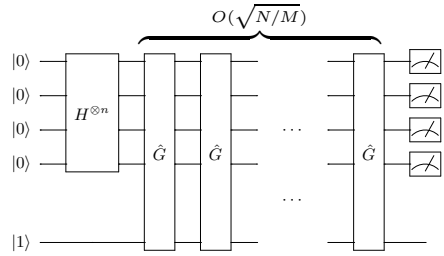
\begin{figure}[h!]
    \centering
$\white{bbbb}\overbrace{\white{bbbbbbbbbbbbbbbbbbbbbbb}}^{O(\sqrt{N/M})}$
\vspace{-2em}
\[
\scalemath{0.7}{
\Qcircuit @C=1.5em @R=0.6em {
   & \lstick{\ket{0}}  & \multigate{3}{H^{\otimes n}} &\multigate{8}{\hat{G}} &  \multigate{8}{\hat{G}} & \qw & & & \multigate{8}{\hat{G}} & \meter\\
   &  \lstick{\ket{0}} & \ghost{H^{\otimes n}} & \ghost{\hat{G}} &  \ghost{\hat{G}} & \qw & & & \ghost{\hat{G}} & \meter\\
   & \lstick{\ket{0}} & \ghost{H^{\otimes n}} & \ghost{\hat{G}} &  \ghost{\hat{G}} & \qw & \push{\cdots} & & \ghost{\hat{G}} & \meter\\
   & \lstick{\ket{0}} & \ghost{H^{\otimes n}} & \ghost{\hat{G}} &  \ghost{\hat{G}} & \qw & & & \ghost{\hat{G}} & \meter\\
   & & & & &  & & & &\\
   & & & & & & \push{\cdots} & & & &\\
   & & & & & & & & &\\
   & & & & & & & & &\\
  & \lstick{\ket{1}} & \qw &  \ghost{\hat{G}} &  \ghost{\hat{G}} & \qw & & & \ghost{\hat{G}} & \qw
   }
   }
   \]
    \caption{Grover's search algorithm}
    \label{fig:groversSearch}
\end{figure}

\begin{figure}[h!]
    \centering
    \[
\scalemath{0.7}{
\Qcircuit @C=1.5em @R=0.6em {
   & \qw & \multigate{8}{\textrm{Oracle}: \ket{x}\ket{y}\rightarrow \ket{x}\ket{f(x)\oplus\ket{y}}} & \multigate{3}{H^{\otimes n}} & \multigate{3}{\hat{M}_0:=2\ket{0}\bra{0}-\id} & \multigate{3}{H^{\otimes n}} & \qw\\
   & \qw & \ghost{\textrm{Oracle}: \ket{x}\ket{y}\rightarrow \ket{x}\ket{f(x)\oplus\ket{y}}} & \ghost{H^{\otimes n}} & \ghost{\hat{M}_0:=2\ket{0}\bra{0}-\id} & \ghost{H^{\otimes n}} & \qw\\
   & \qw & \ghost{\textrm{Oracle}: \ket{x}\ket{y}\rightarrow \ket{x}\ket{f(x)\oplus\ket{y}}} & \ghost{H^{\otimes n}} & \ghost{\hat{M}_0:=2\ket{0}\bra{0}-\id} & \ghost{H^{\otimes n}} & \qw\\
   & \qw & \ghost{\textrm{Oracle}: \ket{x}\ket{y}\rightarrow \ket{x}\ket{f(x)\oplus\ket{y}}} & \ghost{H^{\otimes n}} & \ghost{\hat{M}_0:=2\ket{0}\bra{0}-\id} &  \ghost{H^{\otimes n}} & \qw\\
   & & &\\
   & & &\\
   & & &\\
   & & &\\
   & \qw & \ghost{\textrm{Oracle}: \ket{x}\ket{y}\rightarrow \ket{x}\ket{f(x)\oplus\ket{y}}} & \qw & \qw & \qw & \qw
   }
   }
   \]
    \caption{A single Grover iteration $\hat{G}$}
    \label{fig:singleGroverIteration}
\end{figure}
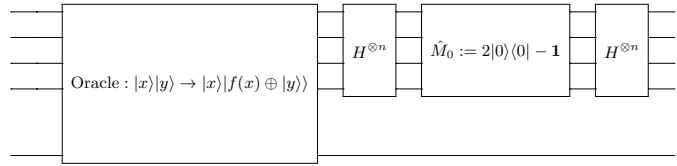

Grover's algorithm \cite{groversAlgo} offers a quadratic advantage in time complexity for search in an unstructured database of $N$ elements. The solution space of size $M$ is defined by the black-box quantum oracle $O_f$ with action $O_f\ket{x}\ket{y}=\ket{x}\ket{q\oplus f(x)}$ where $f(x):\{0,1\}^n\rightarrow \{0,1\}$ is equal to $1$ if and only if $x$ is in the solution space. The time complexity of the algorithm is then $O(\sqrt{N/M})$. As Figure \ref{fig:groversSearch} shows it consists of an iterative application of a Grover step $\hat{G}$ (Figure \ref{fig:singleGroverIteration}) where each Grover step consists of a single application of the Oracle $O_f$ followed by a Grover reflection $H^{\otimes n}(\hat{M}_0:=2\ket{0}\bra{0}-\id)H^{\otimes n}$. To construct a Grover's oracle one needs to find a quantum circuit that implements the oracle $O_f$. Since it is to be reused once for each of the Grover's iterations, its action on the input qubit register $\ket{x}$ must be uncomputed once a conditional flip on the target qubit is performed.

\subsection{BKZ Algorithm}\label{sec:prelimBKZ}
One of the central disciplines in the study of lattices is basis reduction. This consists of iteratively reducing the length of basis vectors, with algorithms taking as input one basis and returning another basis of lattice vectors according to some kind of internal logic. The original basis reduction algorithm is the LLL algorithm \cite{lenstra1982factoring}, but the most commonly used in classical cryptanalysis is the block Korkine Zolotorev (BKZ) algorithm \cite{schnorr1987hierarchy,hanrot2011analyzing,chen2011bkz,bkz20} in which blocks of basis vectors (spanning a sublattice) are reduced with respect to one another, and then the reduction is repeated with a new block of basis vectors.
\begin{definition}
Let $\mathbf{B}=(\mathbf{b}_1, \hdots, \mathbf{b}_n)$ define the lattice $\mathcal{L}(\mathbf{B})$. Let $\mu_{ij} \colon = \langle \mathbf{b}_i, \mathbf{b}_j^* \rangle / \langle \mathbf{b}_j^*, \mathbf{b}_j^* \rangle$. Then $\mathbf{B}$ is \textit{size reduced} if $\mu_{ij} \leq 1/2$  for all $i \geq j$, with $i \leq n$. Then $\mathbf{B}$ is said to be \textit{HKZ reduced} if it is both size reduced and satisfies $\| \mathbf{b}^*_i \| = \lambda_1(\pi_i(\mathcal{L}_{[i,n)}))$.
Furthermore, it is said to be BKZ-$\beta$ reduced if it is 
\begin{itemize}
    \item size-reduced, and
    \item satisfies $\| \mathbf{b}^*_i \| = \lambda_1(\pi_i(\mathcal{L}_{[i, \min(i+\beta-1,n)]}))$,
\end{itemize}
where $\beta$ is the block size.
\end{definition}

The BKZ algorithm can be thought of as an interpolation between the LLL algorithm where neighbouring vectors in the basis are reduced pairwise and then reordered, and HKZ where the block size is $n$. BKZ \cite{bkz20} takes as an input blocksize $\beta$ and searches for shortest vectors in the projected lattices of rank $\leq\beta$. It does this by firstly reducing the lattice spanned by the first $\beta$ vectors, and then the window of vectors moves along by one, until the window reaches the final vector in the basis. Then the window decreases in size until only the final two vectors are reduced. This is known as one `tour', in which all vectors are updated, and $n-1$ calls to the SVP oracle are made. It was shown in \cite{li2020complete} that within $\Theta(\frac{n^2}{\beta^2} \log n)$ tours, the first basis vector is short, with Euclidean length less than 
$$\gamma_\beta^{\frac{n-1}{2(\beta-1)} + \frac{\beta(\beta-2)}{2n(\beta-1)}} \mathsf{vol}(\mathcal{L})^{1/n},$$
where $\gamma_{\beta}$ and $\mathsf{vol}(\mathcal{L})$ are the Hermite constants and the volume of the lattice respectively.

Experiments have suggested two typical uses of BKZ:
\begin{enumerate}
    \item $\beta\approx 20$ for a small $n$ ($ < \approx 80$) or $30\leq \beta \leq 40$ for medium dimension of $n$ ($\approx$ 100). Using these settings BKZ terminates in a reasonable time and tends to improve quality of LLL-reduced basis.
    \item For a high dimension $n$, it is chosen $\beta\geq 40$ and practically the computation needs to be terminated early as the runtime is too long. \cite{bkzTerminated} argues that reasonable early termination still outputs highly reduced bases which is only slightly worse than the full BKZ run. Authors show that if the $\beta$-HKZ reduction (or SVP subroutine) is run $\Omega(\frac{2^3}{\beta^2}(\log n+\log\log\max_i|b_i|))$ times then BKZ returns a basis with a norm $\leq2\nu_{\beta}^{\frac{n-1}{2(\beta-1)}+\frac{3}{2}}$.
\end{enumerate}

The block size $\beta$ parametrizes this interpolation, and larger $\beta$ result in shorter vectors but at higher computation cost.
Currently, using improvements introduced in BKZ2.0, blocksizes of up to dimension 120 are solvable \cite{chen2011bkz}. To subvert lattice security, we would need to perform BKZ with blocksizes of around 400, with Kyber-512 requiring blocksize 406, DiLithium requiring 423 and Falcon-512 needing 411.

\subsection{Hamiltonian approaches to SVP}\label{sec:svpHamiltonian}
Previous works \cite{vqaSVPDavid,vqaSvpMilos,joseph2020not,joseph2022quantum,zhu2022iterative} have investigated quantum variational routines to solve SVP. Central to these works is the definition of a Hamiltonian which encodes the length of lattice vectors into the eigenenergies of the Hamiltonian. Upon measurement of the final state, the configuration that qubits assume corresponds to an eigenstate that can be interpreted as a lattice vector, and the corresponding eigenenergy of that vector is its length. Thus methods that find low energy eigenstates in this context find short length vectors.

The reader wants to know the combination of basis vectors that will give a short vector. So we take the description of a lattice vector $\mathbf{v}=\mathbf{x}\mathbf{B}$ where $\mathbf{x}$ is the coefficient vector, which is read off from the qubit configuration. The Euclidean length of the vector is $\sqrt{\mathbf{v} \mathbf{v}^T}$, but the square root does not affect the ordering of vectors by length, so the authors ignore it. Thus the square of the Euclidean length of a general lattice vector can be written as a combination of the input basis vectors as
\begin{equation}
    \| \mathbf{v} \|^2 = \mathbf{xB} (\mathbf{xB})^T = \mathbf{xBB^Tx^T}
    \label{eq:euclid_len}
\end{equation}
Next the substitution $\mathbf{BB^T}=\mathbf{G}$ is used, where $\mathbf{G}$ is known as the Gram matrix. The aim is to find an optimal $\mathbf{x}$, so the optimization occurs over the coordinates of $(x_1, \hdots, x_n)$.

To turn the expression Eq \ref{eq:euclid_len} into a quantum Hamiltonian, the coordinates $x_i$ are replaced by qudit operators $\hat{Q}^{(i)}$, where a qudit is a $2d$-level quantum system such that when measured, the output is an integer in the range $[-d+1, d]$. Applying this transformation, the Hamiltonian can then be written as
\begin{equation}
    \hat{H} = \sum_{i,j} \hat{Q}^{(i)} \hat{Q}^{(j)} \mathbf{G}_{ij},
    \label{eq:Ham}
\end{equation}
where $\hat{Q}^{(i)}$ represents a qudit operator on qudit $i$, the readout of which is a value for $x_i$. The eigenenergies of the Hamiltonian are the vector lengths squared for all $2^N$ vectors $\mathbf{v}$ which are possible configurations of $N$ qubits. The Hamiltonian can be degenerate, as multiple vectors can have the same length. A notion of successive minima is often used to order the lattice vectors by their length: $\lambda_i(\mathcal{L})$ is the length of the $i^{th}$ shortest non-zero vector of $\mathcal{L}$ while degeneracies are being ignored (i.e. $\lambda_2(\mathcal{L})$ is length of the second shortest non-zero vector in $\mathcal{L}$ no matter how many lattice vectors of length $\lambda_1(\mathcal{L})$ there exist). The goal of the SVP problem is finding a vector in $\mathcal{L}$ of length $\lambda_1(\mathcal{L})$. In order to select the number of qubits required to implement the Hamiltonian of Equation \ref{eq:Ham}, multiple quantities may be utilized. Gaussian heuristics gives an estimate on the length of $\lambda_1(\mathcal{L})$ whereas Minkowski's bound \cite{Kelner2009} bounds this quantity. These information may be used to find a reasonable value of each $d_i$, an integer bound on the coefficienct $|x_i|\leq d_i$. Alternatively, one may choose bounds $d_i$ as to make use of as many qubits as the underlying quantum architecture offers with a hope that short enough vectors still get encoded in the search space. There also exists a more in-depth approach that determines bounds on $|x_i|$ based on reduction of the corresponding dual basis \cite{vqaSvpMilos}.

\section{Figures of merit for resources estimation}\label{sec:figsOfMerit}
In evaluating the implementations we give, we benchmark them with respect to the following figures of merit:

\begin{enumerate}
    \item Space-complexity. The number of qubits that are needed. These can be further divided to ``input'' qubits and ancillary that are all initialized to $\ket{0}$.
    \item Time-complexity. The depth (delay) of the circuit/algorithm is the largest number of 1x1 and 2x2 quantum gates applied sequentially to a qubit
    \item Quantum cost - number of 1x1 and 2x2 reversible gates required to implement a given circuit after a decomposition from higher qubit gates (3 and more) is performed. This is a standard metric and has been used e.g. in \cite{quantumCost}
    \item Number of $T$-gates. Such gates are more expensive when implemented in a fault-tolerant way using because in most used codes are non transverable gates and require magic state distillation.
\end{enumerate}

\section{The oracle circuit design}\label{sec:circuitDesign}

To construct a Grover's oracle for SVP we will first recall how to turn a search problem to a Grover's oracle and then we will outline the steps required for our special case, the SVP. The first thing to do, is to define a Boolean function that returns $1$ for inputs that are solutions to the search and $0$ in all other cases. Then the corresponding function is expressed as a Boolean circuit and then as a reversible circuit, which in its turn is made to a Unitary circuit that acts as the oracle $O_f\ket{x,y}=\ket{x,y\oplus f(x)}$ that uses one extra output register. 

Coming back to SVP, we can break the oracle construction in the following steps. \\
\noindent \textbf{Step 1}: \textit{Define the search space.} Given a basis, one defines the search space as all the lattice vectors that can be produced as linear combinations of the basis vectors, where each of the coefficients $x_i$ of the basis vectors are integers whose absolute values are bounded by a constant $d_i$ (see Section \ref{sec:svpHamiltonian}). If this constant is chosen suitably, the shortest vector of the full lattice is within the vectors spanned in the resulting subspace with a high probability. The authors in \cite{vqaSvpMilos,vqaSVPDavid} demonstrate that $d_i$ scales as $O(\log n)$ for an $n$-dimensional basis. This step gives us the cardinality of the search space (used in Grover's algorithm) which we denote by $N$.\\

A result from \cite{vqaSvpMilos} states that if a lattice vector has length less than $A$, so that $\| x_1 \mathbf{b}_1 + \hdots + x_n \mathbf{b}_n \| \leq A$, then on average it is sufficient to choose $x_i \leq A \cdot \| \hat{\mathbf{b}_i}\|$, where $\hat{\mathbf{b}_i}$ is the $i^{th}$ basis vector of the dual basis $\hat{\mathbf{B}}$. This result translates a bound on the length of the lattice vector into one on the length of the coefficient $x_i$. Bounding the coefficient vectors therefore depends on choice of $A$, and the quality of the dual basis.
The authors describe a preprocessing algorithm which takes in a general lattice basis, and reduces the dual and primal bases. Assuming a search space parameterized by the Gaussian heuristic, they show that an HKZ basis can be found (thus solving SVP) with 
\begin{equation}
    \label{eq:qubit_reqs}
    Q(n) \leq \frac{3}{2}n \log(n) - 2.26n + O(\log n),
\end{equation}
qubits. This means that the search space is $N=2^{c n \log n +o(n)}$ where $c=\frac{3}{2}$. We note that classical enumeration is done in the same search space, and thus would require time complexity of the order of that number.
\\
\noindent \textbf{Step 2}: \textit{Define the solution space.} We do not (a priori) know the length of the shortest vector. However, it is essential to define what is a solution to the search problem before the run of a Grover's algorithm. We define a solution space to be a subset of vectors of the search space (from Step 1) that are smaller than a certain threshold $T$. Finding a suitable $T$ is non-trivial as care needs to be given not to choose $T$ overly small or overly large. In such cases we would either miss the solution to SVP problem or we would converge towards a subset spanned by a large number of lattice vectors, most of which are too large to be considered a valid solution of SVP problem. We could use Minkowski's bound \cite{Kelner2009} to obtain an upper bound on the length of the shortest vector $\lambda_1(\mathcal{L}) \leq \sqrt{n} \cdot D^{1/n}$ where $D$ is the covolume of the of the lattice, equivalent to the modulus of the determinant for any basis. However, being a crude overapproximation, a better approach is to use the Gaussian heuristics. The Gaussian heuristic estimates the number of lattice points inside a given ball $\mathcal{B}$ as $\mathsf{Vol}(\mathcal{B})/\mathsf{Vol}(\mathcal{L})$. It gives a good approximation to the length of the shortest vector as $\lambda_1(\mathcal{L}) \simeq gh(\mathcal{L}) = \sqrt{n/2\pi e} \cdot D^{1/n}$. Choosing $T=\text{gh}(\mathcal{L}))$ thus means that the ball of radius $A$ contains one lattice point for most lattices, which is the shortest vector. Due to the definition of the Gaussian heuristic as $\mathsf{Vol}(B)/\mathsf{Vol}(\mathcal{L})$, the expected number of solutions is $1$, which is necessary to determine the number of iterations for Grover's algorithm. Alternatively, one might use a constant multiple of $\text{gh}(\mathcal{L})$ with constant slightly larger that $1$ to improve a probability of not missing out the shortest lattice vector. The optimal setting is outside scope of this paper and is left as future work.\\

\noindent \textbf{Step 3}: \textit{Find the Boolean a function for this search problem}. This step breaks into two parts. The first part is to evaluate the length of a given vector. Observe that the following equation calculates a resulting squared length of a lattice vector given a row-wise lattice basis matrix $\mathbf{B}$ (of dimension $n\times m$) and a coefficient vector $x=(x_0,...,x_{n-1})$:
$$
\sum_{j=0}^{m-1}\left[\sum_{i=0}^{n-1}x_{i}\mathbf{B}_{ij}\right]^2
$$
As we vary the coefficients $x_i$ within the bounds determined in Step 1, the output register takes on values of squared lengths of all possible lattice vectors inside our pre-defined search space. It is important to note that computing the length is independent of the threshold length for which we consider a vector to be a solution (i.e. on our choices in Step 2), therefore the resources required for a single oracle call do not depend on the choices of multiplicative constant for the Gaussian heuristic. 

Secondly, we perform a projection into two outcomes, separating solutions (below the desired length) from the non-solutions (larger vectors). Again, for this projection, the actual length defining solution, does not change the resource cost.\\
Given an $n\times m$ basis matrix $\mathbf{B}$, the desired quantum oracle implements a function $f: (x_0,\cdots,x_{n-1})\rightarrow \{0,1\}$ defined as
\begin{equation}\label{eq:mappingOracle}
(x_0,\cdots,x_{n-1})\rightarrow\begin{cases}
      1, & \text{if}\  \sum_{j=0}^{m-1}\left[\sum_{i=0}^{n-1}x_{i}\mathbf{B}_{ij}\right]^2\leq T^2 \\
      0, & \text{otherwise}
    \end{cases}
\end{equation}
where $(x_0,\cdots,x_{n-1})\in \mathbb{Z}^n$ are bounded coefficients over which the enumeration is being performed (as determined in Step 1) and $T$ is ideally a tight upper bound on the length of the shortest lattice vector (as determined in Step 2).\\
\noindent \textbf{Step 4}: \textit{Construction of a quantum circuit implementing the oracle.} Given that we know the analytical formula for $f(x)$ found in the previous step, we can now construct a reversible unitary quantum circuit for $O_f$ that acts on the Hilbert space of the input qudits $\ket{x}$ where each of $x_i\in\{x_0,...,x_{n-1}\}$ is a $2d_i$-qudit system as explained in Section \ref{sec:svpHamiltonian} and a single output qubit $\ket{y}$ which gets flipped if and only if $\ket{x}$ refers to a lattice vector belonging to a solution space. See Section \ref{sec:prelimGroversAlgo} for a more detailed explanation on the action of the oracle and Section \ref{sec:encodingIntegerVars} for a further discussion about how to implement the qudits $\ket{x}$ in the given ranges. One can either use standard results from literature about techniques to implement basic arithmetic operations using quantum circuits or one can use a quantum circuit compiler translating an analytical formula into a gate-based quantum circuit. The first option allows for a more flexible tradeoff inbetween different kinds of quantum resources (number of additional ancillary qubits, circuit depth, quantum cost, gate set), essentially tailoring the oracle precisely to a specific underlying quantum architecture. In such a case, one would break the Equation \ref{eq:mappingOracle} into steps of elementary arithmetic operations (as shown in Appendix \ref{app:groversSVPcircDesign}) and consider the optimal way for their implementation. We list some of the state of the art techniques to implement the arithmetic operations needed to build the oracle include in a gate-based quantum circuit model: for addition \cite{Thapliyal2013,li2020efficient}, for multiplication \cite{LiMultiplier,MuozCoreas2017TcountOD,SchonhageStrassen}, for subtraction \cite{adderSubtractor,adder}. The final comparison of the squared length of the vector with a threshold $T^2$ (determined in Step 2) is performed as a subtraction followed by a CNOT gate controlled by a subtraction overflow qubit (indicating that the subtraction result is below zero) and a target qubit $\ket{y}$ which is the output qubit as explained in Section \ref{sec:prelimGroversAlgo}. Alternatively, one might use a quantum circuit compiler that takes as input an analytical equation of the function and outputs a quantum circuit that implements it. Many such software tools exist, e.g. \cite{qiskit,quipper,protoQuipper,scaffc}. Since the oracle will be reused multiple number of times during the execution of Grover's algorithm, it is essential that it uncomputes its action on the input qudit register $\ket{x}$ so that the same register can be used as input to the subsequent runs of the oracle. This can be trivially achieved by applying an inverse gate for each of the gates in a reverse order.\\
\noindent \textbf{Step 5:} \textit{Construction of the Grover's search circuit.} Once a Grover's reflection $H^{\otimes n}\hat{M}_0H^{\otimes n}$ is appended to the oracle $O_f$, a circuit for a single Grover's iteration $\hat{G}$ (Figure \ref{fig:singleGroverIteration}) is constructed. As the Figure \ref{fig:groversSearch} shows, in order to implement the full Grover's search algorithm for the SVP problem, this Grover's iteration needs to be repeated a number of times. This number depends on ratio of the cardinality of the search space $N$ and cardinality of the solution space $M$ and it has been determined in \cite{groversAlgo} that exactly $\lceil\frac{\pi}{4}\rceil\sqrt{N/M}$ iterations are needed. The resulting circuit (prepended by $H^{\otimes n}$ as in Figure \ref{fig:groversSearch}) implements the Grover's algorithm for the SVP problem. It can then by compiled using various software techniques and decomposed into an actual gate set supported by an underlying quantum architecture.

\subsection{Further on Step 3: Encoding the bounded integer enumeration coefficients in terms of qubits}\label{sec:encodingIntegerVars}
We encode each coefficient $x_i$ in a binary representation as shown in \cite{vqaSvpMilos}, \cite{vqaSVPDavid}. Recall, that in Step 1 we chose a bound $|x_i|\leq d_i$ for each of the coefficients. Given $n\times n$ row-wise lattice basis matrix $\mathbf{B}$, the enumeration finds an optimal linear combination of the rows of $\mathbf{B}$ to form the shortest lattice vector. Given the enumeration coefficients $(x_0,...,x_{n-1})$ we express them in a binary expansion
\begin{equation}\label{eq:coeffdecomposition}
    x_i=-d_i+\sum_{j=0}^{\lfloor\log 2d_i\rfloor}c_{ij}2^j
\end{equation}
where each $c_{ij}$ is a newly introduced binary variable. Note that \cite{vqaSvpMilos} proposes also different encoding of the integer coefficients as a technique to avoid trivial solution of the SVP problem (i.e. the zero vector solution) as
$$
x_i=-d_i+\zeta_i d_i+\omega_i(d_i+1)+\sum_{j=0}^{\lfloor\log (a-1)d_i\rfloor-1}c_{ij}2^j
$$
where the enumeration coefficients are $(\zeta_i,\omega_i,c_{i0},c_{i1},...)$ with the advantage that a penalty $L\prod_i\zeta_i$, with a penalty value $L>>0$ can be introduced more efficiently as if the equation (\ref{eq:coeffdecomposition}) is used (in such case the penalty becomes $L\prod_{i,j}c_{ij}$ which is more costly to implement). This different encoding, however, requires more qubits. 
In the approach using variational algorithms \cite{vqaSvpMilos,vqaSVPDavid}, \cite{joseph2022quantum,joseph2020not,zhu2022iterative}, one needs to ensure that the trivial solution does not correspond to the minimal eigenvalue of the corresponding problem Hamiltonian operator, since one performs a minimization and not excluding the zero vector would result in the approach failing to identify the shortest vector by returning the zero vector. In search algorithms, before the final measurement, one obtains a vector that is (with high probability) in the solution subspace, and specifically in equal superposition of the different classical solutions. In our approach to SVP, one defines the solution space as an intersection of the lattice points with a ball placed in the origin with a radius $T$ (determined in Step 2). Therefore, if the ball radius $T$ is optimal, the solution space, even though still containing the trivial solution (and hence increasing the cardinality of solution space by one), contains also sufficiently short lattice vectors (within the ball radius). Generating an equal superposition in a subspace that contains solutions and the zero vector, is not a problem, since by simply repeating the process we can make the probability of obtaining a non-zero vector arbitrarily close to unity. 
Hence after a very few runs of the Grover's algorithm we sample a non-zero short lattice vector within a bound $T$ with a very high probability.

\section{Resource estimation} \label{sec:resourceEstim}

In this Section we analyze both the circuit for the oracle $O_f$ and the circuit for the full Grover's search algorithm. Section \ref{subsec:analyticalEstimation} provides a theoretical analysis to prove scalings for resources required to implement the single oracle $O_f$ and finds polynomial scaling of space complexity (quadratic) and the quantum cost (quintic) in dimensions of the lattice basis. We also prove that time complexity scales linearly. Section \ref{sec:experimentalResourceReqs} then presents experimental results after the circuit for the oracle $O_f$ has been compiled by Quipper \cite{quipper} and plots the experimentally determined scalings of resource requirements. We find that the experimentally determined asymptotes agree with the theoretical analysis and we provide exact numbers for the quantum resources if the circuit is decomposed into Clifford+T gate set. Finally, in Section \ref{sec:solvingSVPusingGrover} using the costs for the oracle we obtain the cost of a single Grover's iteration and subsequently, the overall cost of Grover's search algorithm by repeating the iteration sufficient number of times.

\subsection{Analytical analysis of Oracle resource requirements}\label{subsec:analyticalEstimation}

In this section we present a high-level circuit design. Let $\mathbf{B}$ be an $n\times m$ row-wise matrix of the lattice basis such that the first column is non-negative (this can be achieved by multiplication of some of the rows of $\mathbf{B}$ by $-1$ as needed). This saves a few quantum operations as the sign of the result of multiplication $x_i\mathbf{B}_{i1}$ depends solely on the sign of $x_i$ (see Appendix \ref{app:groversSVPcircDesign} for a more detailed discussion). Recall, that the formula (Equation \ref{eq:mappingOracle} which we are going to denote $\gamma$) to be implemented as the quantum circuit is a function $\gamma:=f: (x_0,\cdots,x_{n-1})\rightarrow \{0,1\}$ defined as
\begin{equation*}
(x_0,\cdots,x_{n-1})\rightarrow\begin{cases}
      1, & \text{if}\  \sum_{j=0}^{m-1}\left[\sum_{i=0}^{n-1}x_{i}\mathbf{B}_{ij}\right]^2\leq T^2 \\
      0, & \text{otherwise}
    \end{cases}
\end{equation*}
We proceed to analyze the scalings of:
\begin{enumerate}
    \item $\text{S}(\gamma)$, a function that returns space-complexity of implementing a quantum circuit corresponding to the function $\gamma$. 
    \item $\text{T}(\gamma)$, a function that returns time-complexity of a quantum circuit implementing the function $\gamma$. 
    \item $\text{C}(\gamma)$, a function that returns quantum cost of a quantum circuit implementing the function $\gamma$. 
\end{enumerate}
These three functions express the figures of merit that we want to consider, and are essentially functions/properties of quantum circuits. We will therefore break these calculations to the steps/sub-circuits that give this function. For notational simplicity, we give as argument of these functions the function/operations that the corresponding circuit implements.

Suppose we use the qudit encoding from Equation \ref{eq:coeffdecomposition} which is a standard, and probably the optimal, choice as discussed in Section \ref{sec:encodingIntegerVars}. We split the calculation of $\gamma$ into subsequent steps which are also depicted in Figure \ref{fig:oracleDiagram}. Note that to the best of our knowledge, all the existing proposals for quantum arithmetic circuits acting on two operands assume that both the operands are of the same size. Suppose for example that we want to implement a sum of two integer quantities $K$ and $L$ encoded in $k$ and $l$ number of qubits respectively with $k\leq l$. Then the first operand needs to be padded with $l-k$ extra qubits initialized to zero to make the encoding of both integer quantities of equal length. Therefore, in this case $\text{S}(K+L)=2\max(k,l)$.
\begin{enumerate}
    \item \textit{Encoding the enumeration coefficients.} By Equation \ref{eq:coeffdecomposition}, $x_i=\sum_{k=0}^{\lfloor\log 2d_i\rfloor}2^kc_k-d_i$ with each $c_k$ being a binary variable (taking values 0 or 1). We implement each $c_k$ with a single logical qubit. Since, in the calculation of $x_i$ a subtraction operation is needed, the number of qubits needed to implement each enumeration coefficient is
    \begin{align*}
        \text{S}(x_i)&=2\max(\lfloor\log 2d_i\rfloor+1, \lfloor\log d_i\rfloor+1)+o(\log d_i)\\
        &=2\lfloor\log 2d_i\rfloor+1+o(\log d_i)\\
        &=\Theta(\log d_i)
    \end{align*}
    where $o(\log d_i)$ accounts for an additional qubit representing a sign of subtraction and additional ancillary qubits depending on a specific subtraction implementation. Because a quantum implementation of subtraction has both linear time complexity and linear quantum cost in the number of input qubits \cite{adderSubtractor},
    \begin{align*}
        \text{T}(x_i)&=\Theta(\text{S}(x_i))=\Theta(\log d_i)\\
        \text{C}(x_i)&=\Theta(\text{S}(x_i))=\Theta(\log d_i)
    \end{align*}
    \item \textit{Multiplication of an enumeration coefficient with an element of a basis vector.} Once an enumeration coefficient $x_i$ is calculated inside the circuit, the next step is to multiply it with an element of a basis vector. Even though $x_i$ needs to be multiplied with each element $\mathbf{B}_{ij}$ for all $0\leq j\leq n-1$, in this step we consider a multiplication for a single value of $j$. An encoding of $\mathbf{B}_{ij}$ requires an additional qubit for $j\geq 1$ (as we assumed $\forall i: \mathbf{B}_{i0}\geq 0$). The space-complexity is therefore
    \begin{align*}
        \text{S}(x_i \mathbf{B}_{ij})&=2\max(Q(x_i), (\lfloor\log \mathbf{B}_{ij}\rfloor+1)+1_{\text{if }j\geq 1})\\
        &=2\max(\theta(\log d_i), \theta(\log \mathbf{B}_{ij}))\\
        &=\Theta(\log \max(d_i,\mathbf{B}_{ij}))
    \end{align*}
    There are many options to choose how to implement the multiplication operation with different asymptotics. \cite{SchonhageStrassen} provides a comparison between state of the art approaches. We continue our analysis with a choice of Schönhage–Strassen algorithm \cite{SchonhageStrassen} as it has in our opinion the best ratio of space complexity to quantum cost (being $O(\log_2^2 n)$ and $O(n \cdot \log n \cdot \log \log n)$ respectively) while the time complexity being relatively small compared to most of other approaches with polynomial time complexities.    
    \begin{align*}
        \text{T}(x_i \mathbf{B}_{ij})&=\Theta(\log^2(\text{S}(x_i \mathbf{B}_{ij})))+\text{T}(x_i)\\
        &=\Theta(\log^2 \log^2 \max(d_i,\mathbf{B}_{ij}))+\Theta(\log d_i)\\
        &=\Theta(\log d_i)\\
        \text{C}(x_i \mathbf{B}_{ij})&=\Theta(\text{S}(x_i \mathbf{B}_{ij}) \cdot \log\text{S}(x_i\mathbf{B}_{ij}) \cdot \log\log \text{S}(x_i \mathbf{B}_{ij}))+\text{C}(x_i)\\
        &=\Theta(\log \max(d_i,\mathbf{B}_{ij}) \cdot \log \log \max(d_i,\mathbf{B}_{ij}) \cdot \log \log \log \max(d_i,\mathbf{B}_{ij}))\\
        &\ \ \ \ +\Theta(\log d_i)\\
        &=\Theta(\log \max(d_i,\mathbf{B}_{ij}) \cdot \log \log \max(d_i,\mathbf{B}_{ij}) \cdot \log \log \log \max(d_i,\mathbf{B}_{ij}))
    \end{align*}
    \item \textit{Multiplication of an enumeration coefficient with a $j$-th column of $\mathbf{B}$.}\label{item:innerSum} We proceed to calculate resource requirements for the inner-sum of $\gamma$, the expression $\sum_{i=0}^{n-1}x_{i}\mathbf{B}_{ij}$. To implement an addition operation taking $n$ operands, the least costly approach is to stack $\Theta(n)$ circuits implementing addition for two operands in in serial-parallel structure resembling a tree with $\Theta(\log n)$ layers. Each layer reduces the number of operands to be summed up by a half (see Figure \ref{fig:oracleDiagram} for a drawing). No other qubits need to be introduced (except possibly some additional ancillary qubits which scale at worst as the scalings of the operands). Because a quantum implementation of addition has both linear time complexity and linear quantum cost in the size of operands \cite{adderSubtractor}. Let $\kappa_j:=\max(\max_i d_i,\max _i \mathbf{B}_{ij})$ be the length of the largest input register out of $\ket{d_0},\ket{d_1},...,\ket{d_{n-1}},\ket{B_{0j}},\ket{B_{1j}},...,\ket{B_{n-1,j}}$. Then
    \begin{align*}
        \text{S}\left(\sum_{i=0}^{n-1}x_{i}\mathbf{B}_{ij}\right)&=n\max_i\text{S}(x_i \mathbf{B}_{ij})\\
        &=\Theta(n\log \max(\max_i d_i,\max _i \mathbf{B}_{ij}))\\
        &=\Theta(n\log \kappa_j)\\
        \text{T}\left(\sum_{i=0}^{n-1}x_{i} \mathbf{B}_{ij}\right)&=\Theta(\log n)\Theta((\text{S}(x_i \mathbf{B}_{ij}))^2)+\text{T}(x_i \mathbf{B}_{ij})\\
        &=\Theta(\log n\times(\log \kappa_j)^2)+\Theta(\log \kappa_j)\\
        &=\Theta(\log n\times(\log \kappa_j)^2)\\
        \text{C}\left(\sum_{i=0}^{n-1}x_{i} \mathbf{B}_{ij}\right)&=\Theta(n)\Theta((\text{S}(c_i \mathbf{B}_{ij}))^2)+\Theta(n\times \text{C}(x_i \mathbf{B}_{ij}))\\
        &=\Theta(n\times(\log \kappa_j)^2)+\Theta(n\times \log \kappa_j\log\log\kappa_j\log\log\log\kappa_j)\\
        &=\Theta(n\times(\log \kappa_j)^2)
    \end{align*}
    \item \textit{Taking square of a resulting vector elements.} To calculate a square of a register $\ket{\psi}$, one can either initialize a new quantum register with the same length and copy the values of logical qubits from $\ket{b}$ ($b$ being a bitstring), essentially creating $\ket{b}\ket{b}$. Note that the copy operation does not conflict with the no-cloning theorem because $\ket{\psi}$ is a register of qubits having values in the computational basis ($0$ or $1$). One can then use a circuit for multiplication to calculate $\ket{b^2}$. Alternatively, one can use a circuit proposed in  \cite{circDesignForTrans} that presents an inverted circuit for the square root operation based on the non-restoring digit recurrence method \cite{qFastPoissonSolver} which in practice lowers time-complexity and quantum cost, although the asymptotes stay the same. We will ignore the cost for the copy operation in the analysis as 1. it would not change the asymptotic scalings and 2. if the latter approach is followed, the copy operation is not performed. So we find that
    \begin{align*}
        \text{S}\left(\left[\sum_{i=0}^{n-1}x_{i} \mathbf{B}_{ij}\right]^2\right)&=2\text{S}\left(\sum_{i=0}^{n-1}x_{i} \mathbf{B}_{ij}\right)\\
        &=\Theta(n\log \kappa_j)\\
        \text{T}\left(\left[\sum_{i=0}^{n-1}x_{i} \mathbf{B}_{ij} \right]^2\right)&=\Theta\left(\log^2\text{S}\sum_{i=0}^{n-1}x_{i} \mathbf{B}_{ij} \right)+T\left(\sum_{i=0}^{n-1}x_{i} \mathbf{B}_{ij} \right)\\
        &=\Theta(\log^2(n\log \kappa_j))+\Theta(\log n\times(\log \kappa_j)^2)\\
        &=\Theta(\log^2(n\log \kappa_j))\\
        \text{C}\left(\left[\sum_{i=0}^{n-1}x_{i} \mathbf{B}_{ij} \right]^2\right)&=\Theta\Biggl(\text{S}\left(\sum_{i=0}^{n-1}x_{i} \mathbf{B}_{ij} \right)\log \text{S}\left(\sum_{i=0}^{n-1}x_{i} \mathbf{B}_{ij} \right)\times\\
        &\ \ \times\log\log \text{S}\left(\sum_{i=0}^{n-1}x_{i} \mathbf{B}_{ij} \right)\Biggr)+\Theta(n)\text{C}\left(\sum_{i=0}^{n-1}x_{i} \mathbf{B}_{ij} \right)\\
        &=\Theta((n\log \kappa_j)\log(n\log \kappa_j)\log\log(n\log \kappa_j))\\
        &\ \ \ \ +\Theta(n)\Theta((n\times\log \kappa_j)^2)\\
        &=\Theta((n\log \kappa_j)\log(n\log \kappa_j)\log\log(n\log \kappa_j))
    \end{align*}
    \item \textit{Calculating a squared length of the resulting vector by performing the outer sum.} The last step to calculate a squared length of the lattice vector corresponding to the enumeration coefficients $x_0,...,x_{n-1}$ is to take the outer sum, i.e. to implement $\sum_{j=0}^{m-1}\left[\sum_{i=0}^{n-1}x_{i} \mathbf{B}_{ij} \right]^2$. We follow the strategy as outlined in Step \ref{item:innerSum} to implement an addition of $m$ operands. Let $\hat{\kappa}:=\max_j\kappa_j$, i.e. it is the length of the largest input register out of $\ket{d_0}$, $\ket{d_1}$, $...$, $\ket{d_{n-1}}$, $\ket{B_{00}}$, $...$, $\ket{B_{0{j-1}}}$, $\ket{B_{1,0}}$, $...$, $\ket{B_{n-1,j-1}}$. The costs of required resources are then
    \begin{align*}
        \text{S}\left(\sum_{j=0}^{m-1}\left[\sum_{i=0}^{n-1}x_{i} \mathbf{B}_{ij} \right]^2\right)&=m\max_{i,j}\text{S}\left(\left[\sum_{i=0}^{n-1}x_{i} \mathbf{B}_{ij} \right]^2\right)\\
        &=m\max_j\Theta(n\log \kappa_j)\\
        &=\Theta(mn\log \hat{\kappa})\\
        \text{T}\left(\sum_{j=0}^{m-1}\left[\sum_{i=0}^{n-1}x_{i} \mathbf{B}_{ij} \right]^2\right)&=\Theta(\log m)\Theta\left(\text{S}\left(\left[\sum_{i=0}^{n-1}x_{i} \mathbf{B}_{ij} \right]^2\right)^2\right)\\
        &\ \ \ \ +\text{T}\left(\left[\sum_{i=0}^{n-1}x_{i} \mathbf{B}_{ij} \right]^2\right)\\
        &=\Theta(\log m)\Theta\left(\left(n\log \kappa_j\right)^2\right)+\Theta(\log^2(n\log \kappa_j))\\
        &=\Theta\left(n^2\log m\times\left(\log \kappa_j\right)^2\right)\\        
        \text{C}\left(\sum_{j=0}^{m-1}\left[\sum_{i=0}^{n-1}x_{i} \mathbf{B}_{ij} \right]^2\right)&=\Theta(m)\Theta\left(\text{S}\left(\left[\sum_{i=0}^{n-1}x_{i} \mathbf{B}_{ij} \right]^2\right)^2\right)\\
        &\ \ \ \ +\Theta\left(m\times\text{C}\left(\left[\sum_{i=0}^{n-1}x_{i} \mathbf{B}_{ij} \right]^2\right)\right)\\
        &=\Theta(m)\Theta\left(\left(n\log \kappa_j\right)^2\right)\\
        &\ \ \ \ +\Theta(m\times (n\log \kappa_j)\log(n\log \kappa_j)\log\log(n\log \kappa_j))\\
        &=\Theta(mn^2\left(\log \kappa_j\right)^2)
    \end{align*}
\item \textit{Deciding whether the resulting vector lies in a solution space}. The last step is to implement the comparison $\sum_{j=0}^{m-1}\left[\sum_{i=0}^{n-1}x_{i} \mathbf{B}_{ij} \right]^2\leq T^2$. For this, we use a subtractor to calculate $\sum_{j=0}^{m-1}\left[\sum_{i=0}^{n-1}x_{i} \mathbf{B}_{ij} \right]^2-T^2$ and we check if the result overflows through zero. Similarly as in Step 1, we need to have the same number of qubit inputs for both operands. Since it is expected that we choose $T^2$ to be lower than the largest vector from our search space, we know that $\text{S}\left(\sum_{j=0}^{m-1}\left[\sum_{i=0}^{n-1}x_{i} \mathbf{B}_{ij} \right]^2\right)\geq\text{S}\left(T^2\right)$ and hence
\begin{align*}
\text{S}\left(\sum_{j=0}^{m-1}\left[\sum_{i=0}^{n-1}x_{i} \mathbf{B}_{ij} \right]^2\leq T^2\right)&=2\text{S}\left(\sum_{j=0}^{m-1}\left[\sum_{i=0}^{n-1}x_{i} \mathbf{B}_{ij} \right]^2\right)\\
&=\Theta(mn\log \hat{\kappa})
\end{align*} We again use a circuit for subtraction as in Step 1 and therefore we have a linear scaling of time-complexity and quantum cost in number of input qubits
\begin{align*}
    \text{T}\left(\sum_{j=0}^{m-1}\left[\sum_{i=0}^{n-1}x_{i}\mathbf{B}_{ij}\right]^2\leq T^2\right)&=\Theta\left(\text{S}\left(\sum_{j=0}^{m-1}\left[\sum_{i=0}^{n-1}x_{i}\mathbf{B}_{ij}\right]^2\right)\right)\\
    &\ \ \ \ +\text{T}\left(\sum_{j=0}^{m-1}\left[\sum_{i=0}^{n-1}x_{i}\mathbf{B}_{ij}\right]\right)\\
    &=\Theta(mn\log \hat{\kappa})+\Theta\left(n^2\log m\times\left(\log \kappa_j\right)^2\right)\\
    &=\Theta(nm+n^2\log\hat{\kappa}+n^2\log m\times\log\hat{\kappa})\\
    \text{C}\left(\sum_{j=0}^{m-1}\left[\sum_{i=0}^{n-1}x_{i}\mathbf{B}_{ij}\right]^2\leq T^2\right)&=\Theta\left(\text{S}\left(\sum_{j=0}^{m-1}\left[\sum_{i=0}^{n-1}x_{i}\mathbf{B}_{ij}\right]^2\right)\right)\\
    &\ \ \ \ +\text{C}\left(\sum_{j=0}^{m-1}\left[\sum_{i=0}^{n-1}x_{i}\mathbf{B}_{ij}\right]^2\right)\\
    &=\Theta(mn\log \hat{\kappa})+\Theta(mn^2\left(\log \kappa_j\right)^2)\\
    &=\Theta(mn^2\left(\log \kappa_j\right)^2)
\end{align*}
\end{enumerate}

Letting $n=m$ (i.e. for the case of square lattices) and taking $\hat{\kappa}=\Theta(\log n)$ (since it grows with the number of qubits dedicated to each enumeration coefficient), we conclude that using the choices for the circuits implementing the arithmetic operations, these would be the resource scalings:

\begin{table}[]
\begin{center}
\addtolength{\tabcolsep}{3pt} 
\renewcommand{\arraystretch}{1.2} 
\begin{tabular}{|l|l|}
\hline
Space Complexity & $\Theta(n^2\log n)$ \\ \hline
Time Complexity &   $\Theta(n^2\log^2 n)$     \\ \hline
Quantum Cost &    $\Theta(n^3\log^2n)$    \\ \hline
\end{tabular}
\addtolength{\tabcolsep}{-3pt} 
\renewcommand{\arraystretch}{1} 
\end{center}
\caption{Calculated asymptotical quantum resource requirements for a single instance of an oracle $O_f$ given a full-rank lattice basis of dimension $n$}
\label{tbl:complexities}
\end{table}

\subsection{Experimental resource requirements scalings determined by classical compilation}\label{sec:experimentalResourceReqs} 

We compiled the circuit for quantum oracle $O_f$ (Equation $\ref{eq:mappingOracle}$) using quipper \cite{quipper} including the uncomputation of any intermediate action performed on the input register. The resource requirements are given at Table \ref{tbl:oracleResources} and are plotted in Figure \ref{fig:quipperOracleResources}. We performed the best-fit of the experimental results to find the following scalings:
\begin{align*}
    \text{Space Complexity} &\approx 3.4n^2\log(n)+97.81n-999.2\\
    \text{Time Complexity}&\approx 0.34n^2\log^2(n)+4028.29\log(n)-5124.38\\
    \text{Quantum\ Cost}&\approx 2.038n^3\log^2(n)+6.3n-2450093\\
        \text{T-Gate count}&\approx 1803.57n^2+2628.43n\log(n)-51449.96n+281006.36
\end{align*}
To find the scaling, \textit{curve\_fit} function from Python's library \textit{scipy} has been used to fit a function of the form $f(n)=an^2\log(n)+bn\log(n)+c\log(n)+dn^2+en+f$ with coefficients $a,b,...,f$ to the obtained experimental data of Space Complexity, $f(n)=an^3\log^2(n)+bn^2\log^2(n)+cn\log^n(n)+dn^3\log(n)+en^2\log(n)+fn\log(n)+gn^3+hn^2+in+j\log(n)^2+k\log(n)+l$ with coefficients $a,b,...,l$ to the obtained experimental data of Quantum Cost and T-Gate Count and a function $f(n)=an^2\log^2(n)+bn\log^2(n)+cn^2\log(n)+dn\log(n)+en^2+fn+g\log(n)^2+h\log(n)+i$ with coefficients $a,b,...,i$ to fit the obtained experimental data of Time Complexity and T-Depth.

\begin{table}[h!]
\centering
\scalebox{0.9}{
\addtolength{\tabcolsep}{3pt} 
\renewcommand{\arraystretch}{1.2} 
\begin{tabular}{l|rrrrrrrrrr}
 & 2 & 5 & 10 & 20 & 30 & 40 & 50 \\
\hline
Space Complexity & 29 & 263 & 1154 & 5165 & 11345 & 23246 & 36056\\
Time Complexity & 344 & 3058 & 7248 & 17352 & 23360 & 38854 & 45440 \\
Quantum Cost & 2006 & 70996 & 467774 & 2817236 & 6311554 & 15910146 & 24831782 \\
Number of T-Gates & 28 & 4718 & 80858 & 542674 & 1258752 & 3232930 & 5097846 \\
T-depth & 62 & 736 & 1220 & 2866 & 3932 & 6336 & 7640 \\
\end{tabular}
\addtolength{\tabcolsep}{-3pt} 
\renewcommand{\arraystretch}{1} 
}
\caption{Oracle resource requirements found by quipper with $\log n$ qubits per each enumeration coefficient.}
\label{tbl:oracleResources}
\end{table}

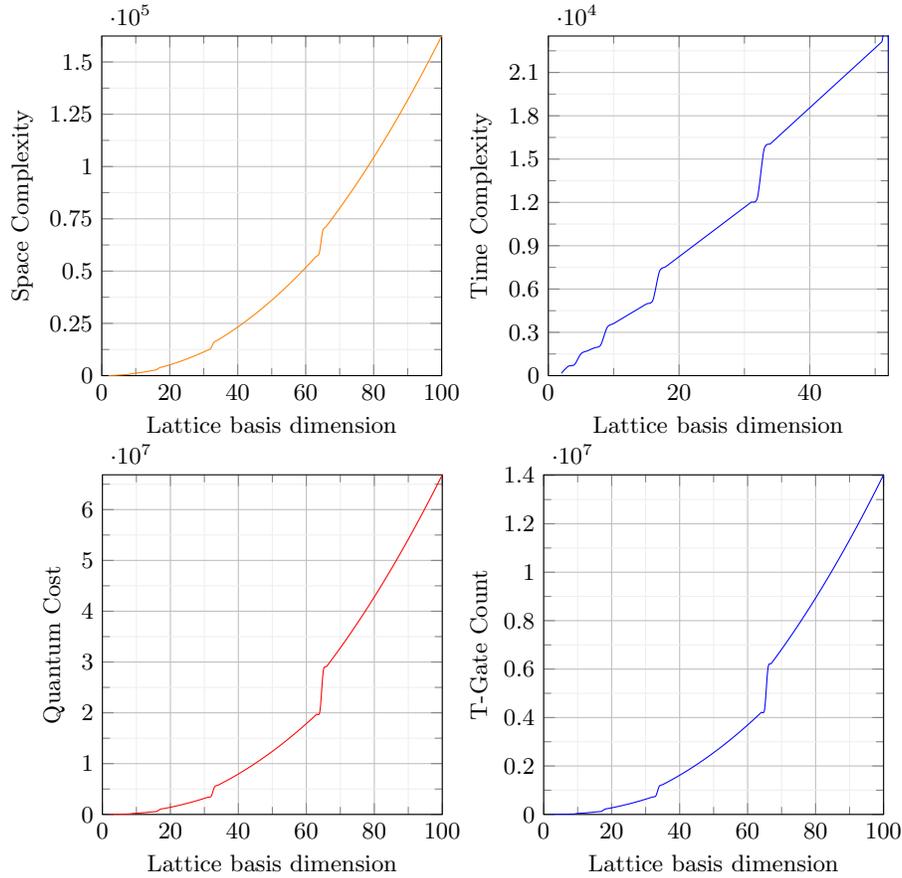
\begin{figure}[h!]
\begin{subfigure}[t]{.5\textwidth}
\begin{tikzpicture}
\begin{axis}[
	xmin = 0, xmax = 100,
	ymin = 0,  ymax = 162407,
	xtick distance = 20,
	ytick distance = 25000,
	grid = both,
	minor tick num = 1,
	major grid style = {lightgray},
	minor grid style = {lightgray!25},
	width = \textwidth,
	height = \textwidth,
	xlabel = {Lattice basis dimension},
	ylabel = {Space Complexity},]

\addplot[
	smooth,
	thin,
	orange,
] file[skip first] {oracle_space_complex.dat};

\end{axis}
\end{tikzpicture} 
\end{subfigure}%
\begin{subfigure}[t]{.5\textwidth}
\begin{tikzpicture}
\begin{axis}[
	xmin = 0, xmax = 52,
	ymin = 0,  ymax = 23525,
	xtick distance = 20,
	ytick distance = 3000,
	grid = both,
	minor tick num = 1,
	major grid style = {lightgray},
	minor grid style = {lightgray!25},
	width = \textwidth,
	height = \textwidth,
	xlabel = {Lattice basis dimension},
	ylabel = {Time Complexity},]

\addplot[
	smooth,
	thin,
	blue,
] file[skip first] {oracle_time_complex.dat};

\end{axis}
\end{tikzpicture}
\end{subfigure}\\
\begin{subfigure}[t]{.5\textwidth}
\hspace{10pt}
\begin{tikzpicture}
\begin{axis}[
	xmin = 0, xmax = 100,
	ymin = 0,  ymax = 66852201,
	xtick distance = 20,
	ytick distance = 10000000,
	grid = both,
	minor tick num = 1,
	major grid style = {lightgray},
	minor grid style = {lightgray!25},
	width = \textwidth,
	height = \textwidth,
	xlabel = {Lattice basis dimension},
	ylabel = {Quantum Cost},]

\addplot[
	smooth,
	red
] file[skip first] {oracle_gc.dat};

\end{axis}
\end{tikzpicture} 
\end{subfigure}%
\begin{subfigure}[t]{.5\textwidth}
\begin{tikzpicture}
\begin{axis}[
	xmin = 0, xmax = 100,
	ymin = 0,  ymax = 14022162,
	xtick distance = 20,
	ytick distance = 2000000,
	grid = both,
	minor tick num = 1,
	major grid style = {lightgray},
	minor grid style = {lightgray!25},
	width =\textwidth,
	height = \textwidth,
	xlabel = {Lattice basis dimension},
	ylabel = {T-Gate Count},]

\addplot[
	smooth,
	thin,
	blue,
] file[skip first] {oracle_tcount.dat};

\end{axis}
\end{tikzpicture}
\end{subfigure}

\caption{Experimental resource requirements for SVP oracle circuit as found by quipper. This plots extended dataset of the data found in Table \ref{tbl:oracleResources}. The ``sudden'' jumps at dimensions being power of 2 (x-axis being ..,32,64,..) are due to rounding in determining number of qubits per enumeration coefficient which is $\lceil\log n\rceil$.}
\label{fig:quipperOracleResources}
\end{figure}

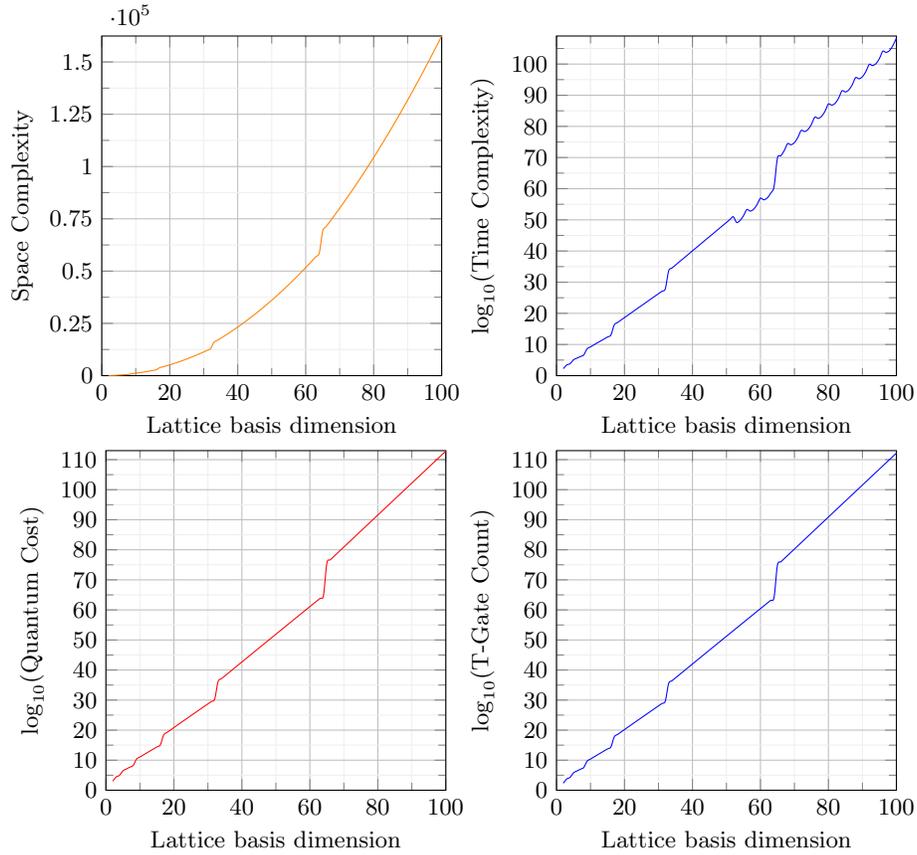
\begin{figure}[h!]
\begin{subfigure}[t]{.5\textwidth}
\begin{tikzpicture}
\begin{axis}[
	xmin = 0, xmax = 100,
	ymin = 0,  ymax = 162407,
	xtick distance = 20,
	ytick distance = 25000,
	grid = both,
	minor tick num = 1,
	major grid style = {lightgray},
	minor grid style = {lightgray!25},
	width = \textwidth,
	height = \textwidth,
	xlabel = {Lattice basis dimension},
	ylabel = {Space Complexity},]

\addplot[
	smooth,
	thin,
	orange,
] file[skip first] {oracle_space_complex.dat};

\end{axis}
\end{tikzpicture} 
\end{subfigure}%
\begin{subfigure}[t]{.5\textwidth}
\begin{tikzpicture}
\begin{axis}[
	xmin = 0, xmax = 100,
	ymin = 0,  ymax=109,
	xtick distance = 20,
	ytick distance = 10,
	grid = both,
	minor tick num = 1,
	major grid style = {lightgray},
	minor grid style = {lightgray!25},
	width = \textwidth,
	height = \textwidth,
	xlabel = {Lattice basis dimension},
	ylabel = {$\log_{10}$(Time Complexity)},]

\addplot[
	smooth,
	thin,
	blue,
] file[skip first] {grover_time_complex.dat};

\end{axis}
\end{tikzpicture}
\end{subfigure}\\
\begin{subfigure}[t]{.5\textwidth}
\hspace{1pt}
\begin{tikzpicture}
\begin{axis}[
	xmin = 0, xmax = 100,
	ymin = 0,  ymax = 113,
	xtick distance = 20,
	ytick distance = 10,
	grid = both,
	minor tick num = 1,
	major grid style = {lightgray},
	minor grid style = {lightgray!25},
	width = \textwidth,
	height = \textwidth,
	xlabel = {Lattice basis dimension},
	ylabel = {$\log_{10}$(Quantum Cost)},]

\addplot[
	smooth,
	red
] file[skip first] {grover_gc.dat};

\end{axis}
\end{tikzpicture} 
\end{subfigure}%
\begin{subfigure}[t]{.5\textwidth}
\begin{tikzpicture}
\begin{axis}[
	xmin = 0, xmax = 100,
	ymin = 0,  ymax = 113,
	xtick distance = 20,
	ytick distance = 10,
	grid = both,
	minor tick num = 1,
	major grid style = {lightgray},
	minor grid style = {lightgray!25},
	width = \textwidth,
	height = \textwidth,
	xlabel = {Lattice basis dimension},
	ylabel = {$\log_{10}$(T-Gate Count)},]

\addplot[
	smooth,
	thin,
	blue,
] file[skip first] {grover_tcount.dat};

\end{axis}
\end{tikzpicture}
\end{subfigure}

\caption{Experimental resource requirements for Grover's search SVP routine as found by quipper. This plots extended dataset of the data found in Table \ref{tbl:groverResources}. Similarly as in Figure \ref{fig:quipperOracleResources}, the ``sudden'' jumps are caused by rounding a function of number of qubits to the nearest higher integer. Note that the Space Complexity remains the same for a single oracle and the Grover's search algorithm: once an oracle is constructed, no additional logical qubits are needed to run the full Grover's search. The graph is plotted again due to reference only.}
\label{fig:quipperOracleResources}
\end{figure}

\subsection{Solving SVP using Grover's algorithm and its resources}\label{sec:solvingSVPusingGrover}

Given the oracle implementation of the previous section, we can use Section \ref{sec:preliminaries} and compute the resources required to find solutions of the SVP for different dimensions. This is done by combining the oracle with other gates to form the Grover iteration and then repeating it sufficient number of times as described in Section \ref{sec:prelimGroversAlgo}. Here we give the break-down of the resources costs for those. When run as part of Grover, both uncomputation (which doubles the depth) and repeated execution $\lceil\frac{\pi}{4}\rceil\times \sqrt{N/M}$ times is required with $N$ being the size of the search space as determined in Step 1 of Section \ref{sec:circuitDesign} and scales as $2^{\frac{3}{2} n \log n +o(n)}$ and $M$ being the size of the solution space as determined in Step 2 of Section \ref{sec:circuitDesign}. Given the integer bounds on the enumeration coefficients $|x_i|\leq d_i$, let $|\text{bin}(d_i)|$ be a minimal number of qubits needed to encode the coefficient in the given range $[-d_i,d_i]$. Then $N=\sum_i |\text{bin}(d_i)|$ and this quantity scales as $O(n\log n)$ as $0\leq i\leq n$ and on average it is enough to assign $\log n$ qubits per coefficient \cite{vqaSvpMilos}. If the bound $T$ in Step 2, Section 4 was chosen optimally, the search space would contain a zero vector, which is always a trivial solution, at least two shortest lattice vectors (as each ball of radius $T$ contains a vectors in pairs which are multiple of $-1$ with each other) and few extra lattice vectors. Hence if $T$ is chosen optimally, the size of the solution space $M\approx 3$ will not be much larger than $3$ and this quantity should ideally stay constant with the increasing lattice dimension. Moreover, we note that by underestimating the value $M$ we are actually requiring more iterations which would result in worse time-complexity. Therefore taking $M=3$ would lower bound the time complexity.
\subsubsection{Extrapolation to ``interesting'' lattice instances}
Classically, the current record of the highest lattice dimension solved as marked by SVP Challenge \footnote{\url{https://www.latticechallenge.org/svp-challenge}} is 186. As discussed in Section \ref{sec:prelimBKZ}, to pose a threat to post-quantum lattice security, one would need to solve SVP problems in lattices of dimensions around 400. Using our extrapolated results from Section \ref{sec:experimentalResourceReqs} we estimate the required quantum resources for 186 and 400 dimensional lattices (Table \ref{tbl:186and400}).
\begin{table}[]
\begin{center}
\addtolength{\tabcolsep}{3pt} 
\renewcommand{\arraystretch}{1.2} 
\begin{tabular}{l|rr}
Lattice Dimension & 186 & 400 \\ \hline
Space Complexity & $6.32\times 10^5$ & $3.3\times 10^6$  \\ 
Time Complexity &  $4.25\times 10^{33}$ & $4.32\times 10^{66}$ \\ 
Quantum Cost &  $5.18\times 10^{36}$ &  $1.03\times 10^{70}$ \\  
T-Gate Count &  $7.09 \times 10^{32}$& $5.98 \times 10^{64}$ \\ 
\end{tabular}
\addtolength{\tabcolsep}{-3pt} 
\renewcommand{\arraystretch}{1} 
\end{center}
\caption{Estimates on quantum resources required to solve 186 and 400 dimensional lattices.}
\label{tbl:186and400}
\end{table}

\begin{table}[h!]
\centering
\scalebox{0.9}{
\stackengine{4pt}{
\addtolength{\tabcolsep}{3pt} 
\renewcommand{\arraystretch}{1.2} 
\begin{tabular}{l|rrrrrrrrrr}
Lattice dimension & 2 & 5 & 10 & 20 & 30\\
\hline
Number of iterations & 1  & 83 & $ 4.75\times10^{5}$ & $ 5.10\times10^{14 }$ & $ 1.71\times10^{22 }$\\
Space complexity & 29 & $2.63\times10^{2 }$ & $1.15\times10^{3}$ & $5.16\times10^{3}$ & $1.13\times10^{4}$ \\
Time complexity & $\hphantom{4}1.72\times10^{2}$ & $ \hphantom{4}1.26\times10^{5 }$ & $ 1.72\times10^{9 }$ & $ 4.42\times10^{18 }$ & $ 2.00\times10^{26 }$  \\ 
Quantum Cost & $ 1.00\times10^{3 }$ & $ 2.94\times10^{6 }$ & $ \hphantom{4}1.11\times10^{11 }$ & $ 7.19\times10^{20 }$ & $ 5.40\times10^{28 }$  \\
Number of T-Gates & $2.1\times10^{2}$ & $6.22\times10^{5}$ & $2.36\times10^{10}$ & $1.53\times10^{20}$ & $1.15\times10^{28}$ \\
\end{tabular}
\addtolength{\tabcolsep}{-3pt} 
\renewcommand{\arraystretch}{1} 
}
{\hspace{0pt}
\addtolength{\tabcolsep}{3pt}
\renewcommand{\arraystretch}{1.2} 
\begin{tabular}{l|rrrrrrrrrr}
Lattice dimension & 40 & 50 & 70 & 90 & 100\\
\hline
Number of iterations     & $ 6.47\times10^{44 }$ & $ 2.56\times10^{73 }$ & $ 3.02\times10^{94 }$ & $ 1.03\times10^{105}$  & $ 6.02\times10^{35 }$\\
Space complexity & $ 2.32\times10^{4  }$ & $3.6\times10^{4}$ & $8.00\times10^{4}$ & $1.31\times10^{5}$ & $1.62\times10^{5}$\\
Time complexity & $ 1.47\times10^{49 }$ & $ 6.15\times10^{74 }$ & $ 8.47\times10^{95 }$ & $ 2.26\times10^{108}$ & $ 1.17\times10^{40 }$\\
Quantum Cost & $ 8.03\times10^{51 }$ & $ 8.40\times10^{80 }$ & $ 1.63\times10^{102 }$ & $ 6.95\times10^{112}$ & $ 4.79\times10^{42 }$\\
Number of T-Gates & $1.02\times10^{42}$ & $1.71\times10^{51}$ & $1.79\times10^{80}$ & $3.50\times10^{101}$ & $1.48\times10^{112}$  \\
\end{tabular}
\addtolength{\tabcolsep}{-3pt} 
\renewcommand{\arraystretch}{1} 
}{U}{l}{F}{F}{S}

}
\caption{Grover resource requirements as found by quipper with $\log n$ qubits per each enumeration coefficient. Calculated by including resource requirements for Grover's reflection to construct a Grover's iteration (Figure \ref{fig:singleGroverIteration}, Section \ref{sec:prelimGroversAlgo}) and multiplying it by number of calculated iterations $\lceil\frac{\pi}{4}\sqrt{N/M}\rceil$.}
\label{tbl:groverResources}
\end{table} 

\section{Grover's improvement over BKZ algorithm}\label{sec:groversImprovement}

We come back to the BKZ algorithm now (see Section \ref{sec:prelimBKZ} for an overview). Recall, that it takes as an input a blocksize $\beta$ and reduces a lattice basis by performing an iterative search for shortest vectors in the projected lattices of rank $\leq\beta$. As mentioned in the preliminaries, there are two main choices for $\beta$ depending on what the lattice dimension is. If it is of a smaller or medium size, one can set $\beta\leq 40$ and terminate in a reasonable time. Otherwise if the lattice dimension is large, $\beta\geq 40$ is usually chosen and the execution of the algorithm is prematurely terminated as it would not be expected to finish in a reasonable time. Further developments, described in BKZ2.0 which incorporate recent algorithmic improvements achieve blocksizes up to 120 \cite{bkz20}.  
For high values of $\beta$, the runtime of BKZ is dominated by the runtime of the SVP oracle. In a single tour, $n-1$ calls to the SVP oracle are made, and many tours may be required, although sometimes the process is halted early as this often produces a good enough length vector.
as shown in \cite{bkzTerminated}.
We suggest two possible approaches to integrating our quantum Grover enumeration algorithm with BKZ, given a lattice basis of dimension $n$:
\begin{enumerate}
    \item Block size $\beta$ remains the same as in the classical \cite{bkz20} scenario, however, each projected block of size $\beta$ would be solved with Grover's algorithm for SVP, described here. Since Grover's algorithm offers quadratic time improvement for the enumeration, the SVP routine would terminate in a shorter time than the classical state of the art. This approach may be necessary in the earliest stages, when the very first large fault tolerant quantum computers are available. But this technique is suboptimal, as the large constants due to Grover imply that the advantages over classical solvers do not appear until a certain `crossover point', which in the medium term is conjectured to be computations which take weeks or months on classical hardware \cite{hoefler2023disentangling}. Thus seeking quadratic speedup on problem sizes already accessible to us classically is unlikely to be the most fruitful approach.

    \item We increase the blocksize $\beta$ that is possible via classical enumeration. While sieving has superior time complexity classically ($2^{0.292\beta + o(\beta)}$, versus $2^{O(\beta \log \beta)}$  for enumeration), it requires exponential space. Increasing the blocksize accessible via enumeration while avoiding the exponential space requirements will make quantum enumeration more appealing, although asymptotically, classical sieving scales better than quantum enumeration. 
    As mentioned in the previous point, the greatest quantum advantages will hail from attacking the largest problem sizes possible in one chunk, thus maximizing blocksize will be the optimal approach long term. For example, if we consider the asymptotic Grover improvement (the factor $1/2$ infront of the $\beta\log\beta$ at the exponent) and ignoring for the moment the extra overhead due to
the implementation, quantumly we could solve blocks of size $\beta\approx 70$ at the same time that classically one would solve blocks of size $40$.    
    
\end{enumerate}

For Level I NIST candidates that are being standardized at the time of writing, blocksizes of over 400 will be necessary to break the underlying hardness assumptions. Despite introducing Grover speedup to the problem, the time complexity of $2^{\frac{c n\log n}{2}+o(n)}$ that we prove is still asymptotically worse than classical sieving, whose main bottleneck is space requirements. We have demonstrated the impact of this Groverized enumeration circuit when integrated into BKZ, the primary approach used to attack lattice cryptosystems, and hence to derive appropriate security parameters. Along with works such as \cite{bai2023concrete} and \cite{bindel2023quantum}, we also conclude that quadratic speed-up quantum algorithms will not pose a significant threat to lattice-based cryptography. In order to properly compare the quantum time complexity we present here with classical algorithms, it is necessary to estimate the speed of implementation of (fault-tolerant) quantum gates which is highly variable depending on the platform in use. Furthermore, the maturity of the technology will set limits upon the blocksize achievable with quantum algorithms such as ours, and thus the overall attack complexity.

\section{Summary and Conclusions}\label{sec:conclusions}

Grover's search algorithm is one of the most important algorithms because of the generality of the speed-up it offers. On the other hand, the generality arises from assuming access to an oracle that identifies the solutions, and any real use of the algorithm would require to implement such oracle without the a-priori knowledge of the solution. In the case of the Shortest Vector Problem, we overcame this difficulty by building on classical enumeration methods and prior Hamiltonian approaches to the SVP. We gave a detailed step-wise description of the oracle construction and then estimated the resources required to implement such an oracle in a quantum device. Incorporating these results to the overall quantum search algorithm of Grover led us to our main results that quantify the resources required to solve SVP in dimension $n$, and we observed that the time complexity scaled as $\Theta(n^2\log^2 n) \times 2^{\frac{3}{4} n\log n+o(n)}$ and space complexity of $\Theta(n^2\log n)$.

Our implementation and resource estimation, while interesting in its own right as an example of a quantum algorithm, does not directly have implications for cryptography -- the main reason to focus on SVP. The reason is that the scaling is still exponential, and classical state-of-the-art methods do not attempt to directly ``brute-force'' and search the full space. 

To stretch the capabilities and explore the impact that Grover's search algorithm can have for SVP we considered using it as a subroutine to a larger, hybrid now, algorithm. Specifically, (quantum) enumeration, is the basis of the BKZ \cite{schnorr1994lattice} algorithm, where one ``brute-forces'' lattices of smaller dimensions and use them to reduce the basis of a larger dimension lattice. We first demonstrated how using larger blocksize, possible due to better time-complexity, can improve the record for solving SVP via the enumeration methods. At the same time, we also highlighted that for dimensions cryptographically relevant ($n\approx 400$ see NIST), our methods are nowhere close to posing a threat.

Finally, to perform a fair comparison with classical methods, one needs to take into account the overheads (and delays) of quantum error correction. When speaking about concrete values, and not asymptotics, these (large) factors are crucial, but depend on many uncotrollable factors\footnote{Error rates, speed of gates (both varying between different hardware), connectivity of qubits, quantum error correcting code and way to perform fault tolerant quantum operations, decoder, etc.} and goes beyond the scope of this paper to compute them.

\section{Acknowledgments}
PW acknowledges support by EPSRC grants EP/T001062/1, EP/X026167/1 and EP/T026715/1, STFC grant ST/W006537/1 and Edinburgh-Rice Strategic Collaboration Awards and  MP acknowledges support
by EPSRC DTP studentship grant EP/T517811/1.

\clearpage
\begin{subappendices}
\renewcommand{\thesection}{\Alph{section}}

\clearpage
\section{Grover's SVP Oracle High-Level Circuit Design}\label{app:groversSVPcircDesign}

Figure \ref{fig:oracleDiagram} shows a high-level design of a quantum circuit implementing the oracle $O_f$ for a full-rank two-dimensional lattice. $B_{ij}$ are elements of the $2\times 2$ lattice basis matrix, $d_i$ are bounds on the enumeration coefficients and all the $c_i$ compose an input register to Oracle $O_f$. The blocks represent arithmetic operations performed as explained step by step in Section \ref{sec:circuitDesign}. The purple lines represent additional qubits which carry information about sign of the binary numbers.
\input{oracleDiagram.tex}

\end{subappendices}
\bibliographystyle{alpha}
\bibliography{bibfile}

\end{document}